\documentclass[12pt]{article}

\usepackage{enumerate,amsmath,graphicx}
\usepackage{amsfonts}
\newtheorem{theorem}{Theorem}

\title{On Finding Maximum Cardinality Subset of Vectors with a
Constraint on Normalized Squared Length of Vectors Sum}

\author{Anton V. Eremeev$^{1,2}$, Alexander V. Kelmanov$^{3,4}$,
Artem V. Pyatkin$^{3,4}$\\ and Igor~A.~Ziegler$^{1,2}$ }
\date{}

\begin{document}

\maketitle

\begin{center}
$^1$ Omsk Branch of Sobolev Institute of Mathematics
SB RAS, Omsk, Russia,\\
$^2$ Omsk State University n.a. F.M.~Dostoevsky, Omsk, Russia,\\
Emails: eremeev@ofim.oscsbras.ru, ziegler.igor@gmail.com\\
$^3$ Sobolev Institute of Mathematics SB RAS,\\
$^4$ Novosibirsk State University, Novosibirsk, Russia,\\
Emails:  \{kelm,artem\}@math.nsc.ru
\end{center}

\abstract{
In this paper, we consider the problem of finding a maximum
cardinality subset of vectors, given a constraint on the
normalized squared length of vectors sum. This problem is closely
related to Problem 1 from (Eremeev, Kel'manov, Pyatkin, 2016). The
main difference consists in swapping the constraint with the
optimization criterion.

We prove that the problem is NP-hard even in terms of finding a
feasible solution. An exact algorithm for solving this problem is
proposed. The algorithm has a pseudo-polynomial time complexity in
the special case of the problem, where the dimension of the space
is bounded from above by a constant and the input data are
integer. A computational experiment is carried out, where the
proposed algorithm is compared to COINBONMIN solver, applied to a
mixed integer quadratic programming formulation of the problem.
The results of the experiment indicate superiority of the proposed
algorithm when the dimension of Euclidean space is low, while the
COINBONMIN has an advantage for larger dimensions.


}

\section{Introduction} \label{sec:intro}
In this paper, we study a discrete extremal problem of searching a
subset of vectors with maximum cardinality, given a constraint on
the normalized squared length of vectors sum. The main goal of the
study is to test experimentally two different approaches to
solving this problem. The first approach is based on the dynamic
programming and the second one is based on the mixed-integer
mathematical programming. We also comment on the computational
complexity of this problem and estimate the time complexity of a
proposed algorithm based on the dynamic programming principles.

The Maximum Cardinality Subset of Vectors with a Constraint on
Normalized Squared Length of Vectors Sum~(MCSV) problem is
formulated as follows.

    \emph{Given}: a set $\mathcal{Y}=\{y_{1},\ldots,y_{N}\}$ of
    points (vectors) from $\mathbb{R}^{q}$ and a number ~$\alpha \in (0, 1)$.

    \emph{Find}:
    a subset $\mathcal{C}\subseteq \mathcal{Y}$ of maximum cardinality such that
    \begin{equation}\label{eqn:criterion}
    \frac{1}{|\mathcal{C}|} \| \sum_{y\in \mathcal{C}} y\|^2 \leq \alpha\frac{1}{|\mathcal{Y}|} \| \sum_{y\in \mathcal{Y}} y\|^2,
    \end{equation}
    where $\|\cdot\|$ denotes the Euclidean norm.

If the given points of the Euclidean space correspond to people so
that the coordinates of points are equal to some characteristics
of these people, then the MCSV~problem may be treated as a problem
of finding a sufficiently balanced group of people of maximum
size.

MCSV problem is closely related to Problem~1 from~\cite{EKP16}.
The main difference consists in swapping the constraint with the
optimization criterion. The problems of finding a subset of
vectors, analogous to the MCSV~problem are typical in the Data
editing and Data cleaning, where one needs to exclude some error
observations from the sample (see
e.g.~\cite{Greco,Osborne,WaaPanSch}). A recent
example of such a problem may be found in~\cite{Ageev2017}, where
a maximum cardinality subset of vectors is sought, given a
constraint that a quadratic spread of points in the subset w.r.t.
its centroid is upper-bounded by a pre-specified portion of the
total quadratic spread of points in the input set w.r.t. the
centroid of that set.

To compare the MCSV to the problem considered
in~\cite{Ageev2017}, we note that
$$
\frac{1}{|\mathcal{C}|} \| \sum_{y\in \mathcal{C}} y\|^2 =
 \sum_{y\in \mathcal{C}} \|y\|^2 - \sum_{y\in \mathcal{C}} \|y - \bar{y}(\mathcal{C})\|^2.
$$
In the right-hand side, the first sum is the total quadratic
spread of points with respect to zero, the second one is relative
to the centroid~$\bar{y}(\mathcal{C})$ of~$\mathcal{C}$. The value
$
    \frac{1}{|\mathcal{Y}|} \| \sum_{y\in \mathcal{Y}} y\|^2 =  \sum_{y\in \mathcal{Y}} \|y\|^2 - \sum_{y\in \mathcal{Y}} \|y - \bar{y}(\mathcal{Y})\|^2
$ characterizes the difference of analogous quadratic spreads in
the initial set. Therefore the MCSV problem asks for a subset of
maximum size such that, in this subset, the two mentioned above
total quadratic spreads differ by not more than $\alpha$ times
from the same difference in the input set $\mathcal{Y}$.

The MCSV problem may be also treated as a Boolean optimization
problem with a quadratic constraint:

\begin{equation}
\label{eqn:qcmcriterion}
  \sum_{i = 1}^{N} x_i \rightarrow \max,
\end{equation}
s.t.
\begin{equation}
\label{eqn:qcmclusterrestriction}
    \sum_{j = 1}^{q}\left(\sum_{i = 1}^{N}y_{i}^{(j)}x_i \right)^2
     \leq
     \alpha\frac{1}{N}   \sum_{j = 1}^{q}\left(\sum_{i = 1}^{N}y_{i}^{(j)}\right)^2
     \cdot \sum_{i = 1}^{N} x_i,
\end{equation}
\begin{equation}
\label{eqn:qcmboolrestriction}
x_i \in \{0,1\}, \quad i = 1,\dots,N,
\end{equation}
 where

$N$ is the cardinality of set $\mathcal{Y}$,

$q$ is the dimension of the Euclidean space,

$y_{i}^{(j)}$ is $j$-th coordinate of the $i$-th vector,

$x_i$ is a Boolean variable, $x_i = 1$ if the $i$-th vector is
included in the solution; otherwise $x_i = 0$ ($i = 1,\dots,N$).

Another problem related to the MCSV is the trading
hubs construction problem, emerging in electricity markets under
locational marginal pricing~\cite{BEGKV,BEGKK,EKP16AIST}. A
trading hub is a subset of nodes of the electricity greed that may
be used to calculate a {\em price index} as an average nodal price
over the hub nodes. This price index may be employed by the market
participants for hedging by the means of futures
contacts~\cite{BEGKV}. Assume that the set of nodes of the
electricity grid which may be included into a hub is
$\{1,\dots,N\}$ and $c_{it}$~is the price at node~$i,$
$i=1,\dots,N,$ at an hour~$t$, $t=1,\dots,T$, where $T$ is the
length of a historic period for which the electricity prices are
observed. Let $p_{rt}$ denote the electricity price of
participant~$r, \ r=1,\dots,R,$ at hour~$t$, and $R$ is the number
of participants. The single hub construction problem consists in
minimizing the sum of squared differences of the prices of
participants from the hub price, requiring that the hub contains
at least~$n_{\rm min}$ nodes:
\begin{equation}\label{eqn:one_hub_crit}
\mbox{Min} \ \sum_{t=1}^T \sum_{r=1}^R (c_t-p_{rt})^2
\end{equation}
s.t.
\begin{equation}
c_t = \frac{\sum_{i=1}^N x_i c_{it}}{\sum_{i=1}^N  x_i}, \
t=1,\dots,T,
\end{equation}
\begin{equation}\label{constr_L}
\sum_{i=1}^N  x_i \ge n_{\rm min},
\end{equation}
 \begin{equation}\label{one_hub_ranges}
  x_i \in \{0,1\}, \ i=1,\dots,N, \ \ c_t \geq 0, \ t=1,\dots,T.
\end{equation}
Here the binary variables $x_{i}$ turn into~1 whenever node~$i$ is
included into the hub. The variables~$c_t$ define the hub price at
time $t, \ t=1,\dots,T$. This problem is proved to be NP-hard
in~\cite{BEGKV}, where a genetic algorithm was proposed for
finding approximate solutions to it. In the special case of a
single market participant problem
(\ref{eqn:one_hub_crit})--(\ref{constr_L}) is equivalent to the
MCSV problem, where the criterion~(\ref{eqn:one_hub_crit}) and
constraint~(\ref{constr_L}) are swapped and instead of a lower
bound on the hub size (which turns into a maximization criterion
in MCSV) we are given an upper bound on the sum of squared
differences of the prices of participants from the hub price. Such
a modification of single hub construction problem may be
appropriate in situations where the required closeness of a hub
price to the prices of participants may be defined, e.g. on the
basis of an observation an already existing hub~\cite{NEPOOL}.

The paper has the following structure. In
Section~\ref{sec:complexity}, we show that MCSV problem is NP-hard
even in terms of finding a feasible solution. An exact algorithm
for solving this problem using the dynamic programming approach is
proposed in Section~\ref{sec:alg}. The algorithm has a
pseudo-polynomial time complexity in the special case of the
problem, where the dimension~$q$ of the space is bounded from
above by a constant and the input data are integer.
Section~\ref{sec:exp} describes how the algorithm is implemented
and a computational experiment is carried out. The purpose of the
experiment is to analyze the algorithm and compare it with the
COINBONMIN solver.

\section{Problem Complexity} \label{sec:complexity}

The following proposition shows that MCSV problem is NP-hard even
in terms of finding a feasible solution.

\begin{theorem}
Finding out whether MCSV has a feasible solution is NP-hard.
\end{theorem}
{\bf Proof.} Consider an instance of the Exact Cover by 3-sets
problem, i.~e. the family of subsets $A_1,\ldots ,A_n$ of a set
$A$ where $|A_i|=3$ for all~$i$ and $|A|=3p$ where $p$~is an
integer. The question is whether there are $p$ subsets in this
family whose union is $A$. This problem is known to be
NP-complete~\cite{GJ}.

Put $q=3p,\ N=n+1$ and for each $i=1,\ldots, n$ let $y_i$ be a
characteristic vector of the set $A_i$ (i.~e. the $j$-th
coordinate of $y_i$ is $1$ if $j\in A_i$ and $0$ otherwise). Put
$y_N=(-1,\ldots, -1)$ and choose $\alpha$ in such a way that
$\alpha  \| \sum_{y\in \mathcal{Y}} y\|^2 < 3$. Then the
constructed instance has a feasible solution if and only if there
is an exact cover by 3-sets. Indeed, if there is no such cover
then for each non-empty set $\mathcal{C}$ we have
$$ \frac{1}{|\mathcal{C}|} \| \sum_{y\in \mathcal{C}} y\|^2 \geq \frac{3}{N} > \frac{\alpha}{|\mathcal{Y}|}  \| \sum_{y\in \mathcal{Y}} y\|^2$$
by the choice of $\alpha$, i.~e. there are no feasible solutions. The opposite implication is trivial.

\section{A Pseudo-Polynomial Time Algorithm for Bounded Dimension of Space} \label{sec:alg}

    In this section, we show that in the case of a fixed dimension $q$ of the space and integer
    coordinates of vectors from~$\mathcal{Y},$ the MCSV problem can be solved in a
    pseudo-polynomial time using the same approach as proposed in~\cite{EKP16}.

    For arbitrary sets~$\mathcal{P}, \mathcal{Q}\subset \mathbb{R}^q$
    define their sum as
\begin{equation}
\label{eqn:rule}
    \mathcal{P}+\mathcal{Q}=
    \{x\in \mathbb{R}^q \ |\  x=y+y', \ y\in \mathcal{P}, \ y'\in \mathcal{Q} \}\,.
\end{equation}
    For every positive integer $r$ denote by
    $\mathcal{B}(r)$ the set of all vectors in $\mathbb{R}^q$ whose
    coordinates are integer and at most~$r$ by absolute value. Then
    $|\mathcal{B}(r)|\le(2r+1)^q.$

    Let $b$ be the maximum absolute value of all coordinates of the input vectors
    $y_1,\dots,y_N$. Our algorithm for the MCSV problem successively computes the subsets
    $\mathcal{S}_k\subseteq \mathcal{B}(bk),\ k=1,\dots,N$,
     where each subset $\mathcal{S}_k$ contains all vectors that can be obtained by
    summing different elements of the  set    $\{y_1,\dots,y_k\}$.

    For $k=1$ we assume $\mathcal{S}_1=\{{\bf 0}, y_1\}$. Then we compute
     \begin{equation}
     \mathcal{S}_k=\mathcal{S}_{k-1}+(\{{\bf 0}\} \cup \{y_k\})
     \end{equation}
     for all
    $k=2,\dots,N$, using the formula~(\ref{eqn:rule}).

    For each element
    $z\in\mathcal{S}_k$ we store an integer parameter~$n_z$ and a subset
    ${\mathcal C}_z\subseteq {\mathcal Y}$ such that
$
    z=\sum_{y\in \mathcal{C}_z} y\,,
$
     where $|\mathcal{C}_z|=n_z$ and $n_z$ is the maximum
    number of addends that were used to produce~$z$.

    When the subset $\mathcal{S}_N$ is computed,
    we find an element $z^*\in\mathcal{S}_N$
    such that $\|z^*\|^2/n_{z^*}\le \alpha\frac{1}{|\mathcal{Y}|} \| \sum_{y\in \mathcal{Y}} y\|^2$
  and the value~$n_{z^*}$ is maximum (if such elements exist in~$\mathcal{S}_N$). The result
  of the algorithm is the
    subset~$\mathcal{C}_{z^*}$ corresponding to the found vector~$z^*$
    or a conclusion that the problem instance is infeasible.
    Let us give a formal outline of the algorithm described above.

    $\mbox{ }$\\
    \noindent {\bf Initialization}\\
    $\mbox{ }$Put ${\mathcal C}_{\textbf{0}} := \emptyset, n_{\textbf{0}} := 0 , {\mathcal C}_{y_1} := \{y_1\}, n_{y_1} :=  1$.\\
    $\mbox{ }$Let~${\mathcal S_1}:=\{\textbf{0}, y_1\}$.\\
    {\bf The main loop:}\\
    $\mbox{ }$ {\bf For} all $k=2,\dots,N$ {\bf do}\\
    $\mbox{ }$ \ \ \ ${\mathcal S_k}:= \mathcal S_{k-1.}$\\
    $\mbox{ }$ \ \ \ {\bf For} all $z \in S_{k-1}$ {\bf do}\\
    $\mbox{ }$ \ \ \ \ \ \ {\bf If} ${\mathcal S_k}$ contains $z'$ such that $z' = z+y_k$ ~{\bf then}\\
    $\mbox{ }$ \ \ \ \ \ \ \ \ \ {\bf If $n_{z'} < n_z +1$} ~{\bf then}\\
    $\mbox{ }$ \ \ \ \ \ \ \ \ \ \ \ \ $n_{z'} =  n_z +1$. \\
    $\mbox{ }$ \ \ \ \ \ \ \ \ \ \ \ \ $\mathcal C_{z'} = {\mathcal C_{z}} \cup \{y_k\}$. \\
    $\mbox{ }$ \ \ \ \ \ \ \ \ \ {\bf End if}.\\
    $\mbox{ }$ \ \ \ \ \ \ {\bf Else}\\
    $\mbox{ }$ \ \ \ \ \ \ \ \ \ ${\mathcal S_k}:={\mathcal S_k} \cup \{z + y_k\}$. \\
    $\mbox{ }$ \ \ \ \ \ \ \ \ \ $n_{z + y_k} :=  n_z +1$. \\
    $\mbox{ }$ \ \ \ \ \ \ \ \ \ $\mathcal C_{z + y_k} := {\mathcal C_{z}} \cup \{y_k\}$. \\
    $\mbox{ }$ \ \ \ \ \ \ {\bf End if}.\\
    $\mbox{ }$ \ \ \ {\bf End for.}\\
    $\mbox{ }$ {\bf End for.}\\
    $\mbox{ }$  Search for $z^*\in S_N$ such that $\frac{\|z^*\|^2}{n_{z^*}}\le \frac{\alpha}{|\mathcal{Y}|} \| \sum_{y\in \mathcal{Y}} y\|^2$
    and~$n_{z^*}$ is maximum.\\
    $\mbox{ }$  {\bf Output} $z^*$ if it exists, otherwise report the problem is infeasible.\\

    Taking into account that computing~$\mathcal{S}_k$ takes
    $\mathcal{O}(q\cdot|\mathcal{S}_{k-1}|)$ operations, we have
    the following
\begin{theorem}
     If the coordinates of the input vectors from $\mathcal{Y}$
     are integer and each of them is at most $b$ by the absolute value
     then MCSV problem is solvable in $\mathcal{O}(qN(2bN+1)^q)$ time.

\end{theorem}

    In the case of fixed dimension $q$ the running time of the
    algorithm\linebreak is~$\mathcal{O}(N(bN)^q)$, i.~e. the problem is
    solvable in pseudo-polynomial time in this special case.

\section{Computational Experiments} \label{sec:exp}

    This section contains the results of testing the
    dynamic programming algorithm~(DP) proposed in
    Section~\ref{sec:alg} and the results of COINBONMIN solver (CBM).
    For the experiments, the DP algorithm was implemented in C++ and
    tested on a computer with Intel Core i7-4700 2.40GHz processor
    and amount of RAM 4GB. First of all, two series of instances were generated
    randomly. To generate these series, we fixed parameter
    $\alpha = 0.1$, the dimension of the space $q = 5$ and number of
    vectors $N = 1000$. In Series~1, the values
    of the vector coordinates varied from -1 to 1, in Series~2
    they varied from -5 to 5 and were integers. In both series
    the coordinates of vectors were generated with uniform
    distribution.

    All testing instances were solved by DP algorithm
and by the package COINBONMIN,
    included in the GAMS package, using the quadratic programming model
    from Section~\ref{sec:intro} (see formulas~(\ref{eqn:qcmcriterion}) to~(\ref{eqn:qcmboolrestriction})).

    \begin{table}
        \caption{CPU time  comparison of the solver CBM and DP Algorithm on Series~1}
        \begin{center}
            \begin{tabular}{ |c|c|c|c|c|c|c|c|c|c|c|c|c|c|c|c| }
                \hline
                Pro-  & CBM & DP & CBM & DP & Pro-  & CBM & DP & CBM & DP\\
         blem & value & value & time & time & blem & value & value & time & time\\
                \hline\rule{0pt}{12pt}
         1 & 977 & 977  &     106,8 & \textbf{40,4} & 16 & 975  & 975  & 91,1 & \textbf{32,1} \\
         2 & 971  & 971  &   131,5 & \textbf{17,0} & 17 & 984  & 984  & 132,3 & \textbf{36,3} \\
         3 & 972  & 972  &   129,3 & \textbf{65,5} & 18 & 986  & 986  & 20,7 & \textbf{18,6} \\
         4 & 986  & 986  & 17,7 & \textbf{15,2} & 19 & 971  & 971  & 180,6 & \textbf{55,1} \\
         5 & 986  & 986  & 18,7 & \textbf{17,2} & 20 & 983  & 983  & 97,5 & \textbf{63,1} \\
         6 & 981  & 981  & 91,7 & \textbf{23,5} & 21 & 978  & 978  & 36,7 & \textbf{27,4} \\
         7 & 984  & 984  & 55,5 & \textbf{19,7} & 22 & 984  & 984  & 191,5 & \textbf{39,1} \\
         8 & 965  & 965  & 232,3 & \textbf{43,2} & 23 & 977  & 977  & 64,6 & \textbf{42,5} \\
         9 & 979  & 979  & 98,6 & \textbf{57,8} & 24 & 958  & 958  & 57,6 & \textbf{31,6} \\
         10 & 968  & 968  & 99,5 & \textbf{20,7} & 25 & 986  & 986  & 20,7 & \textbf{18,6} \\
         11 & 970  & 970  & 127,7 & \textbf{29,1}  & 26 & 966  & 966  & 77,3 & \textbf{40,4} \\
         12 & 990  & 990  & 27,6 & \textbf{21,5} & 27 & 965  & 965  & 324,9 & \textbf{45,3} \\
         13 & 974  & 974  & 25,6 & \textbf{23,1} & 28 & 986  & 986  & 24,9 & \textbf{22,9} \\
         14 & 964  & 964  & 255,1 & \textbf{37,5} & 29 & 965  & 965  & 65,4 & \textbf{47,4} \\
         15 & 981  & 981  & 542,1 & \textbf{46,4} & 30 & 973  & 973  & 408,7 & \textbf{53,8} \\[2pt]
                \hline
            \end{tabular}
        \end{center}
    \end{table}

    \begin{table}
        \caption{CPU time  comparison of the solver CBM and DP Algorithm on Series~2 }
        \begin{center}
            \begin{tabular}{ |c|c|c|c|c|c|c|c|c|c|c|c|c|c|c|c| }
                \hline
                Pro-  & CBM & DP & CBM & DP & Pro-  & CBM & DP & CBM & DP\\
         blem & value & value & time & time & blem & value & value & time & time\\
                \hline\rule{0pt}{12pt}
         1 & 990 & 990  & \textbf{19,1} & 117,3 & 16 & 977 & 977 & \textbf{23,5} & 136,9 \\
         2 & 977 & 977 & 110,3 & \textbf{87,6} & 17 & 990 & 990 & 207,7 & \textbf{99,2} \\
         3 & 985 & 985 & 224,3 & \textbf{93,1} & 18 & 983 & 983 & 289,4 & \textbf{78,2} \\
         4 & 963 & 963 & 199,6 & \textbf{135,4} & 19 & 983 & 983 & \textbf{17,8} & 84,3 \\
         5 & 983  & 983  & \textbf{15,1} & 105,8 & 20 & 968 & 968 & 272,3 & \textbf{95,1} \\
         6 & 982  & 982  & \textbf{62,1} & 96,8 & 21 & 982 & 982 & \textbf{17,9} & 137,1 \\
         7 & 975 & 975 & 527,6 & \textbf{89,2} & 22 & 961 & 961 & 635,6 & \textbf{125,0} \\
         8 & 967 & 967 & 518,3 & \textbf{137,5} & 23 & 984 & 984 & 164,6 & \textbf{106,6} \\
         9 & 979 & 979 & 124,4 & \textbf{94,8} & 24 & 971 & 971 & 167,4 & \textbf{98,6} \\
         10 & 978 & 978 & 112,5 & \textbf{101,4} & 25 & 983 & 983 & \textbf{28} & 84,3 \\
         11 & 965 & 965 & \textbf{65,4} & 126,5 & 26 & 989 & 989 & 203,1 & \textbf{124,2} \\
         12 & 967 & 967 & 127,9 & \textbf{85,7} & 27 & 971 & 971 & 140,6 & \textbf{92,3} \\
         13 & 981 & 981 & \textbf{16,8} & 87,6 & 28 & 987 & 987 & 536,6 & \textbf{116,7} \\
         14 & 974 & 974 & 178,3 & \textbf{81} & 29 & 965 & 965 & \textbf{25,4} & 83,5 \\
         15 & 983 & 983 & 494,2 & \textbf{96} & 30 & 986 & 986 & 66,7 & \textbf{64,8} \\[2pt]
                \hline
            \end{tabular}
        \end{center}
    \end{table}

    The results of the computational experiment for the Series~1
    are presented in Table~1. Here and below, we use the bold font
    to emphasize the best CPU time for each of the instances.
    For all problems of the series, both algorithms have
    found optimal solutions. However in all cases, the DP algorithm found the
    optimal solution faster than CBN. On Series~2,
    in the majority of the cases DP works faster as well
    (The results are presented in Table~2).
    The Wilcoxon signed-rank test showed that the
    CPU times of the DP  and CBN on both Series~1 and Series~2 differ
    with a significance level less than~5\%.

    We also made an experiment, with Series~3, based on the historical
    data on electricity prices from PJM
    Interconnection~(USA), available at\linebreak http://www.pjm.com. The
    dimension of the space turned out to be exceedingly large for
    the DP algorithm to
    meet these challenges, while CBM algorithm was able to solve these problems.
    This is due to a dimension of the space $q=24$. The value of the $\alpha$
    parameter was taken to be 0.1. The results of the experiment
    with Series~3 are shown in
    Table~3. It is worth noting that CBM solver could not find the optimal
    solution to the 4-th instance and managed to find only an approximate solution.

    \begin{table}
        \caption{CPU time  of the solver CBM for electricity prices "PJM Interconnection" }
        \begin{center}
            \begin{tabular}{ |c|c|c|c| }
                \hline
                Problem  & CBM value & CBM time & N \\
                \hline\rule{0pt}{12pt}
                1 & 40* & 0.311 & 43\\
                2 & 118* & 2.293 & 152\\
                3 & 177* & 2.503 & 199\\
                4 & 186 & 87,984 & 199\\
                5 & 223* & 0.867 & 233\\
                6 & 397* & 2,02 & 408\\
                7 & 630* & 3.686 & 642\\
                8 & 625* & 2.776 & 642\\[2pt]
                \hline
            \end{tabular}
        \end{center}
    \end{table}

\begin{figure}[!h]
  \begin{center}
    \includegraphics[width=12cm,height=5cm]{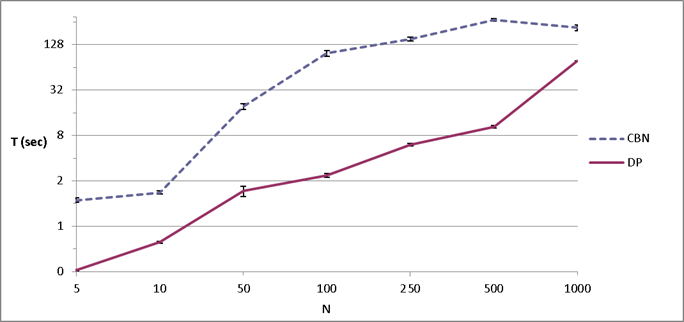}
    \caption{Average CPU time of DP and COINBONMIN as a function of~$N$} \label{fig:nVariation}
  \end{center}
\end{figure}

\begin{figure}[!h]
  \begin{center}
    \includegraphics[width=12cm,height=5cm]{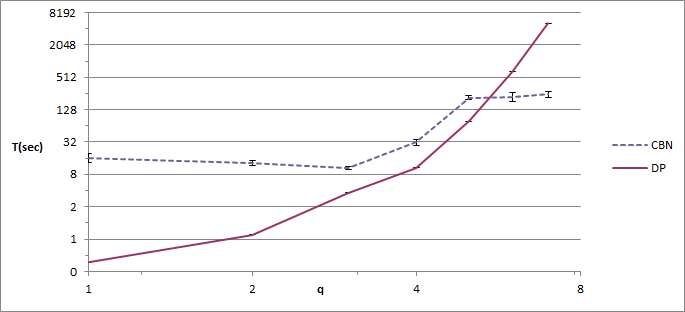}
    \caption{Average CPU time of DP and COINBONMIN as a function of~$q$}
    \label{fig:qVariation}
  \end{center}
\end{figure}

\begin{figure}[!h]
  \begin{center}
    \includegraphics[width=12cm,height=5cm]{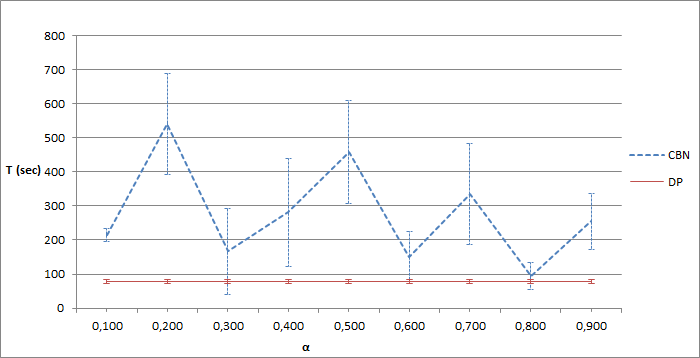}
    \caption{Average CPU time of DP and COINBONMIN as a function of~$\alpha$}
    \label{fig:alphaVariation}
  \end{center}
\end{figure}

    In additional experiments, we generated three series of
    instances in order to
    investigate how the values of $N, q$  and $\alpha$ affect the
    execution time of the DP algorithm and COINBONMIN package. In
    Series~4, we fixed $q = 5$ and $\alpha = 0.1$ and varied $N$
    from 5 to 1000, see Fig.~\ref{fig:nVariation}.
    For Series~5 we put $N = 1000$,
    $\alpha = 0.1$ and varied~$q$ from~1 to~7, see Fig.~\ref{fig:qVariation}.
    In Series~6, we fixed $q = 5$ and $N = 1000$ and varied
    parameter $\alpha$ from~0.1 to~0.9,  see Fig.~\ref{fig:alphaVariation}.
    Six problem instances were randomly generated and solved for each set of
    the parameters mentioned above.
    Average CPU times of both algorithms are presented in
    Figs.~\ref{fig:nVariation}--\ref{fig:alphaVariation}
    where the error intervals show the standard error of the mean.


    The results of experiments with Series~1--5 indicate that in the cases where the dimensionality of
    the space and the maximum value of the coordinates of input vectors
    are not large, the DP algorithm
    is the most appropriate. However, when the dimension of space
    increases, it
    is preferable to use COINBONMIN as a
    mixed integer quadratically constrained program solver (miqcp mode).
    Experiments with Series~6 show that the execution time of COINBONMIN solver is not stable
    w.r.t. variation of~$\alpha$, while the
    CPU time of the DP algorithm does not depend on this
    parameter (which clearly agrees with the DP algorithm description).

\section{Conclusions} \label{sec:conclusion}

 The problem of finding a maximum cardinality subset
of vectors, given a constraint on the normalized squared length of
vectors sum is considered for the first time. It is shown that
even finding a feasible solution to this problem is NP-hard
and an exact dynamic programming algorithm for
solving this problem is proposed. We prove a pseudo-polynomial
time complexity bound for this algorithm in the special case,
where the dimension of the space is bounded from above by a
constant and the input data are integer. An alternative approach
to solving the problem is based on the mixed integer quadratic
programming. Both approaches are compared in a computational
experiment. The results of the experiment indicate that in the
cases where the dimensionality of the space and the maximum value
of the coordinates of input vectors are not large, the dynamic
programming algorithm is the most appropriate. However, when the
dimension of the space increases, it is preferable to use a mixed
integer quadratically constrained program solver, like COINBONMIN.

\bigskip
\subsubsection*{Acknowledgements.}

This research is supported by RFBR, projects 15-01-00462,
16-01-00740 and 15-01-00976.

\bibliographystyle{splncs}

\begin{thebibliography}{9}

\bibitem{Ageev2017}
    Ageev A.A., Kel'manov A.V., Pyatkin A.V., Khamidullin S.A., and
    Shenmaier V.V.: Polynomial approximation algorithm for the
    data editing and data cleaning problem. Pattern Recognition
    and Image Analysis. 27~(3), (2017) (accepted)


\bibitem{BEGKV} Borisovsky P.A., Eremeev A.V., Grinkevich E.B., Klokov
S.A. and Vinnikov A.V.: Trading hubs construction for electricity
markets. In: Kallrath, J., Pardalos, P.M., Rebennack, S., Scheidt,
M. (eds.) Optimization in the Energy Industry. pp.~29--58.
Springer, Berlin, Heidelberg (2009)

\bibitem{BEGKK} Borisovsky P.A., Eremeev A.V., Grinkevich E.B.,
Klokov S.A. and Kosarev N.A.: Trading hubs construction in
electricity markets using evolutionary algorithms. Pattern
Recognition and Image Analysis. 24~(2), 270--282 (2014)


\bibitem{EKP16AIST} Eremeev A.V., Kel'manov A.V. and  Pyatkin A.V.:
On complexity of searching a subset of vectors with shortest
average under a cardinality restriction. In: Proc. of
International Conference on Analysis of Images, Social Networks
and Texts (AIST'2016), Yekaterinburg, Russia, April 7-9, 2016,
pp.~51--57. Springer, Cham (2017)

\bibitem{EKP16} Eremeev A.V., Kel'manov A.V. and  Pyatkin A.V.:
 On the complexity
of some Euclidean optimal summing problems. Doklady Mathematics.
93~(3), 286--288 (2016)

\bibitem{GJ} {Garey M.R. and Johnson D.S.}: Computers and intractability.
A guide to the theory of NP-completeness. W.H.~Freeman and
Company, San Francisco (1979)

\bibitem{Greco} Greco, L. Robust Methods for Data Reduction
Chapman and Hall/CRC (2015)

\bibitem{NEPOOL} NEPOOL Energy Market Hub White Paper and
Proposal. Hub Analysis Working Group NEPOOL Markets Committee
(2003)

\bibitem{Osborne} Osborne J.W. Best Practices in Data Cleaning: A
Complete Guide to Everything You Need to Do Before and After
Collecting Your Data. 1st Edition. SAGE Publication, Inc. Los
Angeles (2013)

\bibitem{WaaPanSch} de Waal T., Pannekoek J., Scholtus S. Handbook of
Statistical Data Editing and Imputation. John Wiley and Sons, Inc.
Hoboken, New Jersey (2011)

\end{thebibliography}

\end{document}